\documentclass[12pt]{article}
\usepackage{epsfig}

\newcommand{\ed}{\end{document}}
\newcommand{\be}{\begin{equation}}
\newcommand{\ee}{\end{equation}}
\newcommand{\ba}{\begin{eqnarray}}
\newcommand{\ea}{\end{eqnarray}}
\newcommand{\baz}{\begin{eqnarray*}}
\newcommand{\eaz}{\end{eqnarray*}}
\newcommand{\bb}{\begin{thebibliography}}
\newcommand{\eb}{\end{thebibliography}}

\begin{document}

\mbox{}

\begin{center}
{\large\bf   Nonlocality of the Nucleon Axial Charge
and Solar Neutrino Problem }

\vskip 0.5cm

 N.I. Kochelev  

\vspace{5mm}
{\small\it
 Bogolyubov Laboratory of Theoretical Physics,\\
Joint Institute for Nuclear Research,\\
Dubna, Moscow region, 141980 Russia}

\vspace*{1cm}

\end{center}
\begin{abstract}
It is shown that the specific space distribution of the nucleon axial
charge may lead to the significant reduction of the cross section
of the reaction $p+p\rightarrow D+e^++\nu$ and may provide
 the mechanism  to  explain
observed  deficit of solar neutrinos.

\end{abstract}

\vspace{0.5cm}

The deficit of solar neutrinos is one of the longstanding unsolved problem
in astrophysics (see review \cite{bahcall}).  
The reaction $p+p\rightarrow D+e^++\nu$ is most important 
process which cross section determines the solar neutrino flux from pp cycle.
Its  value has never been  directly measured and only theoretical
predictions exist so far. There is a wide spread opinion that 
theoretical uncertainties
in the calculation are very small and therefore one can conclude that it is 
impossible to explain the small observed flux of solar neutrinos
 without involving
new physics (e.g. neutrino oscillations etc).

In this Letter we will show that rather extended space   
distribution of nucleon 
axial charge  may lead to significant reduction of the cross section of
the reaction  $p+p\rightarrow D+e^++\nu$.

The main assumption which was used to calculate the rate of this reaction
was the hypothesis of the locality of nucleon axial charge  \cite{bahcall1}.
As the result, the cross section is proportional to product of
two factors
\be
\sigma_{FA}^{p+p\rightarrow D+e^++\nu}
\propto g_A^2|\int \Psi_D^*\Psi_{pp}d\vec{r}|^2,
\label{section}
\ee
where factor $g_A$, axial charge of nucleon, comes from weak interaction matrix element, 
and second factor is overlapping of  $\Psi_D$, 
deuteron wave function and $\Psi_{pp}$, the initial wave 
function of two protons.
We should emphasize that such approach can be  correct only  for 
the point-like
distribution of the nucleon axial charge, i.e. only for the case when
  the characteristic scale of
axial charge distribution is much  smaller than the corresponding 
scale related to overlapping of 
 initial and final hadron wave functions.

Recently Jaffe has shown \cite{jaffe} that the distribution of axial charge
of the nucleon is rather nonlocal  and in chiral limit 
$m_\pi\rightarrow 0$   one third of $g_A$ comes from infinite distance 
$R\rightarrow \infty $ from the center of nucleon. It means that in 
this limit one should expect
that cross section of the reaction 
$p+p\rightarrow D+e^++\nu$ should be much smaller than its factorized
value (\ref{section})
\be
\sigma^{p+p\rightarrow D+e^++\nu}(m_\pi\rightarrow 0)=
\frac{4}{9}\sigma_{FA}^{p+p\rightarrow D+e^++\nu}.
\label{red}
\ee
The chiral limit is a rather good approximation for many calculations 
in hadronic 
physics and therefore it would be difficult to believe that 
factorization hypothesis 
would give the correct answer in the case of the physical pion mass. 
\begin{figure}[htb]
\centering
\epsfig{file=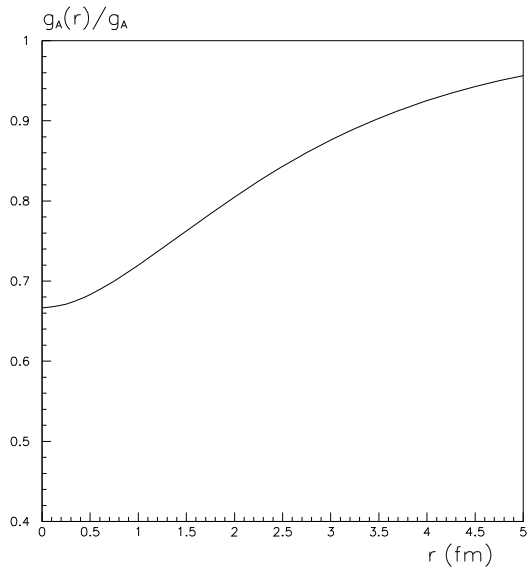,width=8cm}
\vskip 0.5cm
\caption{\it The contribution to nucleon  axial charge 
from distances $r<R$. }
\end{figure}
For finite mass of pion the contribution to axial charge from the  distances
 $r<R$ is \cite{jaffe}
\be
g_A(R)=g_A(1-\frac{1}{3}(1+m_\pi R)e^{-m_\pi R}),
\label{charge}
\ee
where $g_A=1.26$ is total nucleon axial charge.  
In Fig.1 the distribution of $g_A(R)$ is shown.
It is evident that the distribution is quite extended and therefore
the locality approximation which was used to get Eq.(\ref{section})
is not very well justified. 

Nonlocality of axial charge leads to suppression 
of the reaction  $p+p\rightarrow D+e^++\nu$ due to smaller 
value of effective $\tilde{g}_A $ which should enter in Eq. 
(\ref{section}).

One can  roughly   estimates  the effect of suppression 
by using the formula 
\be
\frac{\sigma^{p+p\rightarrow D+e^++\nu}(m_\pi)}{\sigma_{FA}^{p+p\rightarrow D+e^++\nu}}
\approx \frac{g_A^2(R_{eff})}{g_A^2}.
\label{suppr}
\ee
In Eq.(\ref{suppr}) $R_{eff}$ is the effective size of interaction 
region which should
be of the order of a characteristic size of  deutron  
$r_m\approx 2 fm$, defined as the rms-half distance between two nucleons
\be
r_m^2=\frac{1}{4}\int_0^\infty (u^2(r)+w^2(r))r^2dr,
\label{rms}
\ee
with $u(r)$ and $w(r)$ are $S$ and $D$ waves of deuteron.
\cite{deutron}
\footnote{ More accurate estimation will be published elsewhere.}.
Therefore we get the large reduction factor of 
\be
\frac{\sigma^{p+p\rightarrow D+e^++\nu}(m_\pi)}{\sigma_{FA}^{p+p\rightarrow D+e^++\nu}}
\approx 0.65.
\label{ratio}
\ee

Thus, it is shown that nonlocality  of the nucleon axial  charge 
may lead to the large decreasing of the rate of the basic solar reaction 
for neutrino production and may give the mechanism to 
  explain  the  solar neutrino  deficiency.

The author  is grateful to  A.E.Dorokhov, V.A. Bednyakov, S.B. Gerasimov,
E.A.Kuraev and V.Vento 
for 
useful discussion.
This work is  partially  supported by  RFBR-01-02-16431
and INTAS-2000-366 grants.

\end{document}